\documentstyle[twocolumn,prl,aps]{revtex}
\begin{document}
\title{ Comment on ``High-Temperature Series Analysis of the 2D 
		Random-Bond Ising Ferromagnet"}
\maketitle
{In a recent Letter\cite{ROD}, Roder {\it et al} 
claimed that their high temperature series analysis
of the two dimensional random bond Ising ferromagnet (RBIF)
conclusively supported the prediction by Shalaev, Shankar, 
and Ludwig (SSL)\cite{SHA}.
The claim is based on the observation that
their ``ln-Pade" analysis of the magnetic susceptibility ($\chi$)
assuming 
\begin{equation}
\chi \sim  t^{-7/4} |\ln t|^{p}, \label{eq:ssl}
\end{equation}
yields estimate of the logarithmic exponent $p$ that is 
consistent with the predicted value of SSL, $p=7/8$.

Being a perturbation theory, the theory of SSL is supposed to be
more correct as the degree of disorder becomes smaller,
that is, as the value of $J_2/J_1$ becomes closer to one for the
2D RBIF.
When the strength of disorder is extremely small, on the other hand,
the critical behavior of the disordered Ising system must be almost 
indistinguishable from that of the pure system, so 
asymptotic expression of SSL, Eq.\ref{eq:ssl} for $\chi$,
is supposed to hold for extremely narrow scaling regime only. 
The remaining regime is uneffected by the presence of the disorder
and maintains the scaling behavior of the pure system.
Thus, in the context of the theory 
there generally exists a crossover from the critical behavior
of the pure system to Eq.\ref{eq:ssl} as $t \to 0$, which is reflected
in the expression
\begin{equation}
\chi \sim  t^{-7/4} \left [ 1 + g |\ln t| \right ]^{\gamma^{\prime}},
\label{eq:eff_ssl}
\end{equation}
with the value of the logarithmic exponent $\gamma^{\prime}=7/8$.

Note that Eq.\ref{eq:eff_ssl} reduces to the asymptotic form
Eq.\ref{eq:ssl} only when $t$ is extremely small or the
value of $g$ is extremely large.
The value of the $g$ is supposed to increase smoothly with $J_2/J_1$
from $g=0$ at $J_2/J_1=1$, but the theory  
is not able to determine $g$ as a function of $J_2/J_1$.
Since the crossover temperature is a priori unknown,
Eq.\ref{eq:eff_ssl} instead of Eq.\ref{eq:ssl}
should be used for the analysis of series 
expansion or of Monte Carlo (MC) data.

The authors in the Letter simply assume that for $5 \le J_2/J_1 \le 10$
the value of $g$ is sufficiently large and that their series 
safely represents the asymptotic regime of SSL. 
Their estimate of $p\simeq 7/8$ from ln-Pade analysis, however, 
indicates that the value of $\gamma^{\prime}$
be larger than the predicted value of SSL.
This can be easily seen from Fig.2 of the Letter;  
for example for $J_2/J_1=3$ where the prediction of SSL is supposed 
to be more correct than for $J_2/J_1 \ge 5$, 
their estimated value of $p$ is just 0.3 whereas the value of 
$\gamma^{\prime}$ is supposed to be 7/8.

The apparent plateau in the estimate of $p$ (Fig.2 of the Letter)
is surprising in light of the monotonically increasing 
critical exponent with $J_2/J_1$ when analyzed 
assuming pure power law critical behavior  (Fig.1): 
It is an elementary mathematical fact that $\chi \sim t^{-\gamma}$
with $\gamma (> 7/4)$ increasing with $J_2/J_1$ is approximated
with increasing value of $p$ in Eq.\ref{eq:ssl}
rather than its fluctuating values. 
The monotonic increment of the critical
exponent was clearly observed in the previous 
MC studies as well\cite{KIM_SITE,KIM_BOND}. 
In fact MC study on the RBIF\cite{KIM_BOND} and the
series analysis yield completely agreeing estimates of the 
critical exponents. 
With this agreement 
the plateau in the value of $p$ for the wide range 
of $5 \le J_2/J_1 \le 10$ is rather spurious, 
probably resulting from subjective choice 
of the different orders of series terms 
for the different values of $J_2/J_1$ in their analysis.


Their remarks about previous MC studies 
are also mainly incorrect. Especially, previous MC study\cite{KIM_SITE}
that supported varying critical exponent with the strength of
random disorder is not based on finite size scaling analysis but
on the careful measurements of the thermodynamic values  of 
various physical quantities.
It was shown that the thermodynamic data of correlation 
length and the magnetic susceptibility fit equally well to the 
scenario of varying critical exponent and to the predictions of SSL.
However,  the data of the specific heat at least for 
strongly disordered case was manifestly inconsistent with the
double logarithmic behavior predicted by SSL. 

To sum up:
Estimate of the logarithmic exponent is very sensitive depending on
which critical singularity between Eq.\ref{eq:ssl} and Eq.\ref{eq:eff_ssl} 
is used for analysis, and it cannot be justified
that the value of $g$ is so large 
for $J_2/J_1\ge 5$ that Eq.\ref{eq:ssl} is a valid one 
for their analysis.
In general, $p$ must be regarded as a lower bound of $\gamma^{\prime}$  
so that $p \simeq 7/8$ actually indicates $\gamma^{\prime} \gtrsim 7/8$. 
Furthermore, the plateau in the value of $p$ seems to be spurious. 
We thus conclude that their claim is groundless.
In parallel with MC study, it would be interesting to see the
series analysis of the specific heat for strongly disordered case.
\par
\medskip\noindent
Jae-Kwon Kim \par
School of physics, Korea Institute for Advanced Study,\\
207-43 Cheongryangri-dong, 
Seoul  130-012, Korea
}


\begin{references}
\bibitem{ROD} A. Roder {\it et al}, Phys. Rev. Lett. {\bf 80},
		4697 (1998).
\bibitem{SHA} B.N. Shalaev, Sov. Phys. Solid State {\bf 26}, 1811 (1984);
              R. Shankar, Phys. Rev. Lett. {\bf 58}, 2466 (1987);
              A. W. W. Ludwig, Phys. Rev. Lett {\bf 61}, 2388 (1988).
\bibitem{KIM_SITE} J.-K. Kim and A. Patrascioiu, Phys. Rev. Lett {\bf 72},
              2785 (1994); Phys. Rev. B {\bf 49}, 15764 (1994).
\bibitem{KIM_BOND} J.-K. Kim, cond-mat/9502053. 
\end{references}
\end{document}